\providecommand{\PathR}{./}    %
\providecommand{\PathS}{./}

\makeatletter							
\def\input@path{{\PathR}}
\makeatother

\RequirePackage[l2tabu, orthodox]{nag}	
\documentclass[
final, 
reprint,
groupedaddress,
runinaddress,
showpacs,showkeys,preprintnumbers,
nobibnotes,
longbibliography,
amsmath,amssymb,
aps,
pra,
floatfix,
]{revtex4-1}

\renewcommand
\setcitestyle{super,numbers} 

\usepackage{graphicx}
\usepackage{dcolumn}
\usepackage{bm}
\newcommand{\diff}{\ensuremath{\operatorname{d}\!}}

\usepackage{standalone}				
\usepackage{import}					
\usepackage{tikz}					%
\usepackage{pgfplots}				%
\pgfplotsset{compat=newest}		    %

\begin{document}
	
	\preprint{APS/123-QED}
	
	\title{Lattice Boltzmann Method for Heterogeneous Multi-class Traffic Flow}
	
	\author{Romain \textsc{No\"{e}l}}
	\email{romain.noel@emse.fr}
	\author{Laurent \textsc{Navarro}}%
	\email{navarro@emse.fr}
	\affiliation{%
		Mines Saint-\'{E}tienne, Univ. Lyon, Univ. Jean Monnet,\\ INSERM, U 1059 Sainbiose, Centre CIS, F - 42023 Saint-\'{E}tienne France
	}%
	\author{Guy \textsc{Courbebaisse}}%
	\email{guy.courbebaisse@creatis.insa-lyon.fr}
	\affiliation{%
		Univ. Lyon, INSA-Lyon, Universit\'{e} Claude Bernard Lyon 1, UJM Saint-\'{E}tienne, CNRS, INSERM,
		CREATIS UMR 5220, U1206, F69621, Lyon, France
	}%
	
	\date{\today}
	
	\begin{abstract}
		The traffic modelling often keeps the mesoscopic scale in the theoretical sphere because the integro-differential nature of its equations. In the present work we suggest to use the lattice Boltzmann method to overcome these difficulties. In particular, the method has a strong theoretical foundation. An improved version of the lattice Boltzmann method for multi-class and heterogeneities, has been introduced here. Its ability to reproduce the fundamental diagram is proved here, for both single-class and multi-class flows. This allows easily simulating complex and realistic cases of mixture of multi-class traffic flow. These simulations are able to capture jamming in various traffic situations such as road merging, reduction of number of lanes or change of speed limits.
	\end{abstract}
		
	\maketitle
	
	
	\section{Introduction}
	
	Since the end of the twentieth century, with the increase of personal car and road jam, traffic modelling has become a subject in the centre of economic, ecological and social considerations. In this context, constructors and road operators look forward to having a better understanding and anticipation of the traffic flows, in order to promulgate the optimal practical solutions.
	
	In this perspective, the models should handle the heterogeneous characteristics of the traffic flow. One of its major heterogeneity is its composition by multiple classes (or categories) of vehicles. Road operators and car drivers agree on the role of lorries. Due to their longer dimensions and heavier weight, they are incriminated for faster deterioration of the structures, densification of traffic situations, slowing of flows and faster creation of traffic jams for a longer time.
	
	Models have to be able to capture various nonlinear phenomena, such as those responsible for the growth of traffic jams.
	
	Commonly traffic models are classified in three groups: microscopic, mesoscopic and macroscopic.
	First, microscopic models describe the behaviour of each driver individually. So, the interaction between two vehicles can be studied finely. This level of detail allows including the psychological aspect of the drivers. At the microscopic scale, the lane-changing and the overtaking questions are crucial. These models are very adapted to simulate the evolution of the traffic inside cities' streets but also what could be the new traffic behaviours with intelligent or unmanned vehicles. The widely used method in these approaches are the cellular automaton~\cite{Chopard_1996, Nagel_1992, Schadschneider_1993, Nagel_1998} and the car-following theory~\cite{Pipes_1967, Newell_2002, Newell_1993a}. The main counterpart of the microscopic scale is the numerical resources needed to simulate large areas or numerous vehicles. This is due to the number of differential equations to solve with the car-following and the repeat simulations in order to improve signal-to-noise ratio with the cellular automaton.
	
	At the opposite of the spectra, the macroscopic models have high computational efficiency. This efficiency is mainly obtained thanks to the reduced number of macroscopic variables describing the traffic flow and the nature of the partial differential equations ruling these variables~\cite{Lighthill_1955, Richards_1956, Payne_1971}. Such models are the mixture results of the empirical behaviour and hydrodynamics equations. It is an appropriate model for long roads where the position and velocity of each vehicle are not at stake nor significant.
	
	Mesoscopic models use a statistical description to recover the macroscopic equations with a finer level of detail. These approaches adapt the kinetic theory of gases ideas to traffic situations~\cite{Prigogine_1971}. They are based on Boltzmann-like equations, therefore they lie on the microscopic interactions, which give "physical" explanation and foundations to the resulting behaviours. On the one hand, the main mesoscopic variable is the distribution function, reducing the numbers of variables and the computation time. On the other hand, as a price to pay to have fine details and numerical efficiency, is a harder work in functional spaces plus the complexity of the Boltzmann-like equations and their solutions. 
	
	The lattice Boltzmann method (LBM) is born after the statistical averaging for lattice gas cellular automaton, in order to have meaningful results when simple simulations are subjected to numerical noise. But the LBM has a strong theoretical base lying on the Boltzmann equation~	\cite{He_1997a}. Thus, it seems to be a natural framework to build numerical schemas for mesoscopic models.
	
	The multi-class effects are challenging and have been studied at the macroscopic scale~\cite{Logghe_2008, Jiang_2004}. Nevertheless, there are fewer studies at a microscopic scale with cellular automaton~\cite{Lan_2005} or with car-following~\cite{Hidas_2002, Peeta_2005}. The available work for multi-class traffic flow at mesoscopic scale is mainly theoretical~\cite{Hoogendoorn_2000, Hoogendoorn_2001a, Chanut_2005, Shvetsov_1999}.
	
	In the present work, a statistical description of continuous kinetic model is introduced, before to discuss the effects of multi-class heterogeneity on continuum models. Then the construction of the lattice Boltzmann methods and the formulations associated, for both the single class and heterogeneous multi-class frameworks. This is followed by numerical simulations in order to validate the suggested model. Finally, discussions and conclusions are performed in the last section.

	\section{Statistical Description}
	\subsection{Continuous Kinetic Models}
	To build a continuous mesoscopic and kinetic model, one would study the density of vehicles $\rho$ at a spatial position $x$ along a highway, at a time $t$. Obviously, on multi-lane roads, several vehicles can be at the same time and position with a different velocity. This remark is also true, with single lane road, since the representative elementary volume (REV) used for the mesoscopic scale is such that it contains many vehicles. 
	Naturally, this leads to introduce $f(x,\xi,t)$ the distribution of vehicles with the velocity $\xi$. By working in the phase space, trivially, the density, average speed $v$ and flow $q$ are given by:
	\begin{align}
	\rho(x,t)&=\int f(x,\xi,t)\diff\xi \\
	q(x,t)&=\rho(x,t) v(x,t) =\int \xi f(x,\xi,t)\diff\xi.
	\end{align}
	In order to describe the evolution of any distribution of particles, Boltzmann suggested using the sum of a transport term with interactions between these particles. Since the vehicle can be considered as particles, the Boltzmann equation seems perfectly adapted to vehicle transport problems linked to traffic flow. 
	Thus, Prigogine, Herman and Andrews~\cite{Prigogine_1960, Prigogine_1971}, suggested describing the evolution with no "external forces" and by decomposing the collision-interaction operator into two terms. The \emph{Prigogine-Boltzmann equation} for traffic flow is
	\begin{equation}
	\left(\frac{\partial}{\partial t}+ \xi\frac{\partial}{\partial x}\right) f= \Omega(f,f) =\left(\frac{\partial f}{\partial t}\right)_{rel} + \left(\frac{\partial f}{\partial t}\right)_{int}
	\end{equation}
	where the left-hand side member is the transport term, $\Omega$ is the collision-interaction operator, $\left(\frac{\partial f}{\partial t}\right)_{rel}$ and $\left(\frac{\partial f}{\partial t}\right)_{int}$ are respectively the relaxation term and the interaction term.
	
	The relaxation term is the reckoning that drivers will toward a certain speed called desired velocity. This desired velocity for all drivers is described by a distribution function $f^d$, and it is reached after a certain time $\tau$ called the relaxation time.
	Therefore, the relaxation term can be expressed as:
	\begin{equation}
	\left(\frac{\partial f}{\partial t}\right)_{rel}=-\frac{f(x,\xi,t)-f^d(x,\xi,t)}{\tau}.
	\end{equation}
	
	The interaction term renders the overtaking process. When a fast incoming vehicle attains a slower one, the slow one is not affected but if the fast one cannot overtake it has to slow down in a short period. The interaction term can be given by:
	\begin{equation}
	\left(\frac{\partial f}{\partial t}\right)_{int}=(1-p)f(x,\xi,t)\int(\xi'-\xi)f(x,\xi',t)\diff\xi'
	\end{equation}
	where $p$ denotes the average probability to overtake slow vehicles.
	
	This model has been criticised by some authors~\cite{Munjal_1969}. Paveri-Fontana highlighted some problems with this model when $\tau$ and $p$ are constant. He also suggested a close enhanced model~\cite{Paveri-Fontana_1975}. Moreover, he showed that Prigogine-Boltzmann equation and the Paveri-Fontana improved Boltzmann equation yield same macroscopic equation up to the second order.
	
	Thus, as a first approach the Prigogine-Boltzmann equation is a simpler and good description of the traffic flow if $\tau$ and $p$ are linked to the vehicle density. A common relationship between the parameter $p$ and the density $\rho$ is $\gamma=\tau\rho(1-p)$, where $\gamma$ is a constant. In order to have a model closer to the conventional lattice Boltzmann method, the Prigogine-Boltzmann equation can be rewritten
	\begin{equation}
	\left(\frac{\partial}{\partial t}+ \xi\frac{\partial}{\partial x}\right) f(x,\xi,t)=
	\frac{f^{(0)}-f}{\tau'}
	\end{equation}
	with
	\begin{align*}
	f^{(0)}=& \frac{\tau'}{\tau}f^{d}+\left(1-\frac{\tau'}{\tau}\right)\rho(\xi-v) \\
	\tau'=& \frac{\tau}{1+\gamma}.
	\end{align*}
	In the following, the distribution $f^{(0)}$ is called equilibrium distribution function. 
	
	Some authors continued this path to create generic kinetic traffic model adapted to many situations~\cite{Helbing_2001}. Despite the discussions about the validity of their assumptions (specially about the vehicular chaos), the continuous kinetic models seem to be good tools to deal with traffic flows.

	\subsection{Continuum Kinetic Multi-Class Traffic Approach}
	Some authors~\cite{Hoogendoorn_2000, Costeseque_2015} constructed continuum kinetic model for multi-class flows, through numerous inter-class interaction. However, some authors~\cite{Chanut_2005, Bagnerini_2003} reminded that the use of multi-class distribution is easier.
	
	In this work, an alternative formulation is used. Indeed, guided by the intuition that every driver interacts in the same manner with all the other vehicles without independently of the class of the other vehicles. Therefore, the construction of the equilibrium distribution for each class should be the function of the global density over the class and the local density of its own class to respect the continuity equation. 
	
	Let us denote with the subscript $c$ one class of vehicle out of $N_c$ the number of classes. Thus, the mesoscopic multi-class Boltzmann-like equations are
	\begin{align}
	\left(\frac{\partial}{\partial t}+ \xi\frac{\partial}{\partial x}\right) f_c(x,\xi,t)=&
	\frac{f_c^{(0)}(\rho, \rho_c)-f_c}{\tau_c'} \\
	\rho_c(x,t)=& \int f_c(x,\xi,t)\diff\xi \\
	q_c(x,t)=\rho_c v_c(x,t)=&\int \xi f_c(x,\xi,t)\diff\xi \\
	\rho=\sum_{c=1}^{N_c}\rho_c \qquad
	\qquad q=&\rho v=\sum_{c=1}^{N_c}q_c. \label{eq_sum_rho_g}
	\end{align}
	
	To be exact, densities $\rho$ and $\rho_c$ considered here should be interpreted as non-dimensional densities or equivalently as occupation rates. Indeed, let us consider the case where there is two classes, the first one is made of vehicles of length $L_1$ and respectively the second of length $L_2$. If the length of the representative elementary volume (REV) is $D_x$, therefore the maximum number of vehicles from the first class at the position $x$ is denoted $\overline{\rho_1}(x)$ and has the maximum value of $D_x/L_1$. This value is reached when there is no free space between vehicles, which leads to the definition of $\rho_c$ as the occupation rate. Thus for any class, the relationship between $\rho$ and $\overline{\rho}$ is 
	\begin{equation}
	\overline{\rho_c}(x)=\frac{D_x}{L_c}\rho_c(x).
	\end{equation}
	
	Moreover, the global density can only be expressed in terms of vehicles number equivalent to vehicles of one class. In the previous example, the number of vehicles from the second class equivalent to the ones of the first class is $\overline{\rho_{2/1}}=\overline{\rho_{2}}L_2/L_1$. And so, if $\overline{\rho_{tot/1}}$ is the total number of vehicles equivalent to those of the first class, this number is given by
	\begin{equation}
	\overline{\rho_{tot/1}} = \sum_{c=1}^{N_c}\overline{\rho_{c}}\frac{L_c}{L_1} = \rho\frac{D_x}{L_1}
	\label{eq_dim}
	\end{equation}
	which is more convenient to compute with rate occupations (see equation (\ref{eq_sum_rho_g})).
	
	The major drawback with kinetic models (single or multi-class) is the integro-differential nature of the equations. This fact, combined with the lack of physical properties, is responsible for absence of analytical solutions. This is why numerical methods are for now necessary.

	\section{Lattice Boltzmann Method}
	\subsection{Lattice Boltzmann Model for Heterogeneous Traffic Flows}
	
	The lattice Boltzmann Method (LBM) is a discretisation of the continuous Boltzmann equation. This discretisation is performed on all three space, time and phase space. The time discretisation gives an explicit schema. The phase space discretisation is often characterised by the appellation \emph{DnQm} where $n$ specifies the physical dimension of the problem and $m$ is the number of points used to condense the phase space. The denomination lattice is linked to the regular spatial discretisation and the \emph{DnQm} schema used to connect the spatial points; commonly the points of the phase-space are chosen to coincide with the spatial ones~\cite{Succi_2001}.
	
	Commonly in the LBM approach, the assumption of a collision-interaction operator composed only of a relaxation term is made. This assumption is named BGK in reference to Bhatnagar Gross Krook~\cite{Bhatnagar_1954}.
	Thus, under the BGK approximation the lattice Boltzmann equation (LBE) reads~\cite{Wolf-Gladrow_2000, Chopard_2002}
	\begin{equation}
	f_{i}({ {x}}+{ {e}}_{i}\delta _{t},t+\delta _{t})=f_{i}({ {x}},t)+{\frac {1}{\tau}}(f_{i}^{(0)}-f_{i})
	\end{equation}
	where $f_i$ is the distribution evaluated in $e_i$ the discrete velocities (associate to the phase space), $\delta_t$ is the time step, therefore the space or lattice step is given by $\delta_x/\delta_t=e$.
	
\begin{figure}
		\def\svgwidth{0.99\linewidth}
		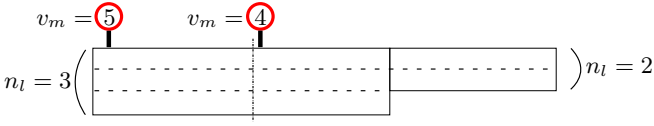
		\caption{Schematic of a road with changing of speed limit and number of lane.}
		\label{road_schema}
	\end{figure}

	For the numerical resolution the LBE is divided in two steps. The first step is the collision-interaction step defined by the equation
	\begin{equation}
	f_{i}
	({ {x}},t+\delta _{t})=f_{i}({ {x}},t)+{\frac {1}{\tau}}(f_{i}^{(0)}-f_{i}).
	\end{equation}
	This is usually followed by the streaming step defined by:
	\begin{equation}
	f_{i}({ {x}}+{ {e}}_{i}\delta _{t},t+\delta _{t})= f_{i}
	({ {x}},t+\delta _{t}).
	\label{stream}
	\end{equation}
	Nevertheless, the last equation must be adapted to take into account possible lane number changes (see fig.~\ref{road_schema}). Therefore, to face these possibilities, we suggest to turn the eq. (\ref{stream}) into: 
	\begin{equation}
	f_{i}({ {x}}+{ {e}}_{i}\delta _{t},t+\delta _{t})= f_{i}
	({ {x}},t+\delta _{t})\frac{n_l(x)}{n_l(x+{e}_{i}\delta _{t})}
	\label{stream2}
	\end{equation}
	with $n_l$ is the number of lanes for a given spatial point.
	
	The macroscopic variables are recovered by using the classic following system:
	\begin{align}
	\rho(x,t)&= \sum_{i=0}^{m-1} f_i(x,t) \\
	q(x,t) &= \rho v(x,t)= \sum_{i=0}^{m-1} e_i f_i(x,t).
	\end{align}

	The LBM in its classical form can solve various problems related to transport of particles, under certain conditions like the compressibility or the viscosity. 
	The equilibrium distribution function has to be in adequacy with the physical phenomena underlying that one desires to capture.
	This distribution function is in the Euler conservation case uniquely found through the application of mathematical theorem~\cite{Cercignani_1988} and physical conservation laws.		
	
\begin{figure}[h]
		\def\svgwidth{0.99\linewidth}
		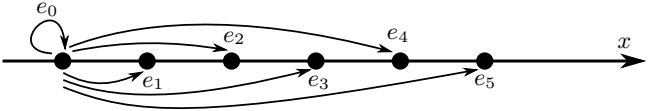
		\caption{Schematic of an asymmetric $D1Q6$ network.}
		\label{D1Q6_schema}
	\end{figure}

	When applying the LBM for road traffic~\cite{Meng_2008, Shi_2016}, the lack of conservation laws makes the appreciation of the equilibrium distribution function harder. As some authors suggested~\cite{Meng_2008} a 	
	shrewd workaround is to construct it from the observable data. The other noticeable difference with the LBM applied to traffic problems is the phase space. It is $\mathbb{R}^n$ (symmetric) for fluids or gases but it becomes $\mathbb{R}^{+}$ (asymmetric) with roads since the dimension of roads is one and vehicles almost never use reverse gear for something else than to park (see fig.~\ref{D1Q6_schema}). 
	
	Previous studies~\cite{Meng_2008} suggested choosing an equilibrium distribution function that offers interesting capacity to model empirical phenomena. We introduced here a completeness of this empirical model to deal with heterogeneities, allows defining the equilibrium distribution function with
	\begin{align}
	f^{(0)}_i=& 
	\begin{cases}
	\frac{\displaystyle \rho(x)}
	{\displaystyle 1+\sum_{i=1}^{v_m} e_i^2\exp\left(-\frac{e_i^2\widetilde{\rho}(x)}{1-\widetilde{\rho}(x)}\right)} 
	& \text{for } i=0 \\
	\frac{\displaystyle e_i^2 \exp\left(-\frac{e_i^2\widetilde{\rho}(x)}{1-\widetilde{\rho}(x)}\right)\rho(x)}
	{\displaystyle 1+\sum_{i=1}^{v_m} e_i^2 \exp\left(-\frac{e_i^2\widetilde{\rho}(x)}{1-\widetilde{\rho}(x)}\right)}
	&	\forall\,i\in[\![ 1,v_m]\!] \\
	\qquad \qquad 0 \hfill\forall\,i\in&\hspace{-6pt}[\![ v_m+1,m-1]\!]				
	\end{cases} \\
	&\widetilde{\rho}(x)=\frac{\displaystyle \sum_{i=0}^{v_m} \rho(x+e_i)}{v_m +1}
	\end{align}
	where $v_m$ is the maximum speed of vehicles (it can be the speed limit, like on fig.~\ref{road_schema}, or higher values, if ones wants to capture over-speeding) and $\widetilde{\rho}$ is the most important parameter. It can be seen as the reachable forward occupation rate. In other words, it is the density, that the drivers feel in front of them, in which they will have to navigate.
	Obviously, the maximum speed of vehicles $v_m$ is an integer lower than the maximum speed used to model the system $m-1$. This also means, that the variations of $v_m$ along the road can only be multiple of the lattice step $\delta_x/\delta_t=e$. 
	
	This equilibrium distribution function expressed the fact that drivers under a constant relaxation time, change their desired speed with the variation of the forward reachable occupation rate. It is a manner to express that they are intelligent enough to adapt their speed with the traffic.
	
	This model has to be completed with "virtual boundary condition" to prevent having density higher than one when too many vehicles from different cells want to reach the same cell. Naturally, the vehicles from the further cells have to slow down.
	
	\begin{widetext}
		\begin{align}
		B_v(f_{i}(x))=&\begin{cases}
		f_{i}^{*}(x-e_i\delta_t)&=\begin{cases}
		0 & \text{if }
		\displaystyle\rho_{test}-\sum_{j=i+1}^{m}f_j(x-e_j\delta_t)\frac{n_l(x-{e}_{j}\delta _{t})}{n_l(x)}>1 \\
		f_{i}(x-e_i\delta_t) \phantom{ \ + f_{i-1}(x-e_i\delta_t)} & \text{else}
		\end{cases} \\
		f_{i-1}^{*}(x-e_i\delta_t)&=\begin{cases}
		f_{i}(x-e_i\delta_t)+ f_{i-1}(x-e_i\delta_t) & \text{if }
		\displaystyle\rho_{test}-\sum_{j=i+1}^{m}f_j(x-e_j\delta_t)\frac{n_l(x-{e}_{j}\delta _{t})}{n_l(x)}>1 \\
		f_{i-1}(x-e_i\delta_t) & \text{else}
		\end{cases}
		\end{cases} \label{eq_virt_bound}\\
		{\rho_{test}}(x)=&\displaystyle \sum_{i=0}^{m} f_{i}(x-e_i\delta_t)\frac{n_l(x-{e}_{i}\delta _{t})}{n_l(x)}. \label{eq_rho_test}
		\end{align}
	\end{widetext}
	
	A virtual boundary condition inspired by~\cite{Meng_2008}, can be expressed through the function $B_v$ (see eq.~(\ref{eq_virt_bound})), in which the quantity $\rho_{test}$ (see eq.~(\ref{eq_rho_test})) represents the density that would happen if all the drivers could stream as they wish, i.e. if the density could be greater than one and so that car crash could happen. 
	
	To avoid any confusion, since in the definition of $B_v$ the comparison is made with 1, the definition of $\rho$ or $\rho_{test}$ used here is the occupation rate varying from 0 to 1.
	
	Thus, we suggest to rewrite the streaming step that takes into account the virtual boundaries:
	\begin{equation}
	f_{i}({ {x}}+{ {e}}_{i}\delta _{t},t+\delta _{t})= B_v(f_{i}
	({ {x}},t+\delta _{t}) ).
	\label{stream_VB}
	\end{equation}

	\subsection{Lattice Boltzmann Model for Multi-Class Traffic}
	Strengthened by its impressive results to model the Navier-Stokes equations, researchers quickly tackled more complex problems like mixture of fluids. A noticeable work has been done in case of immiscible mixture~\cite{Shan_1993}. But despite heavy machinery vehicles tend to form a continuous lane on highways, the second lane is most of the time full of personal car. Therefore, the hypothesis of immiscibility seems not to be the most relevant in a first approach.
	
	At the same time, some pioneers initiate the work on miscible fluids~\cite{Holme_1992}, improved some years after by adding the thermodynamics~\cite{Inamuro_2002}. 
	
	Even if, few publications deal with the use of cellular automatons for multi-class traffic flows~\cite{Ez-Zahraouy_2004}; the use of the lattice-Boltzmann method to solve heterogeneous multi-class traffics has never been studied. 
	
	In order to take into account the necessity for the equilibrium density function to depend on the global density and the class density, and since the variable $\widetilde{\rho}$ can be understood as the density felt forward drivers; it seems natural for a first approach to link the dependency to the global density to $\widetilde{\rho}$. 
	Thus, the macroscopic variable related to the eq.~(\ref{eq_sum_rho_g}) can be written as
	\begin{align}
	\rho_c=\sum_{i=0}^{m-1}f_{c,i}(x,t)  \qquad q_c=\sum_{i=0}^{m-1}e_i f_{c,i}(x,t) \\
	\rho=\sum_{c=1}^{N_c}\rho_c=\sum_{i=0}^{m-1}\sum_{c=1}^{N_c}f_{c,i}(x,t) \\
	q=\sum_{c=1}^{N_c}q_c=\sum_{c=1}^{N_c}\sum_{i=0}^{m-1}e_i f_{c,i}(x,t).
	\end{align}
	
	Then, the respect of conservation equations leads to a new formulation of the 
	equilibrium density function:
	\begin{align}
	f^{(0)}_{c,i}=& 
	\begin{cases}
	\frac{\displaystyle \rho_c(x)}
	{\displaystyle 1+\sum_{i=1}^{v_{m,c}} e_i^2\exp\left(-\frac{e_i^2\widetilde{\rho_c}(x)}{1-\widetilde{\rho_c}(x)}\right)} 
	& \text{for } i=0 \\
	\frac{\displaystyle e_i^2 \exp\left(-\frac{e_i^2\widetilde{\rho_c}(x)}{1-\widetilde{\rho_c}(x)}\right)\rho_c(x)}
	{\displaystyle 1+\sum_{i=1}^{v_{m,c}} e_i^2 \exp\left(-\frac{e_i^2\widetilde{\rho_c}(x)}{1-\widetilde{\rho_c}(x)}\right)}
	&	\forall\,i\in[\![ 1,v_{c,m}]\!] \\
	\qquad \qquad 0\hfill\forall\,i\in&\hspace{-6pt}[\![ v_{c,m}+1,m]\!]			
	\end{cases} \\
	&\widetilde{\rho_c}(x)=\frac{\displaystyle \sum_{i=0}^{v_{m,c}} \rho(x+e_i)}{v_{m,c} +1}
	\end{align}
	where $v_{m,c}$ is the maximum speed of the class $c$ vehicles, and $\widetilde{\rho_c}$ is the reachable forward occupation rate for the class $c$.
	
	The virtual boundary conditions also have to follow the same logic. One should notes that since the $B_v$ function changes the values of the distribution at a point backward from a given point where variables are evaluated, its algorithmic application should be backward recursive. The variable $\rho_{test}$ has to be the reflection of the global density, while the modifications over densities has to be accomplished per class. Therefore, the virtual boundaries can be expressed by:
	\begin{widetext}
		\begin{align}
		B_v(f_{i,c})=&\begin{cases}
		f_{i,c}^{*}(x-e_i\delta_t)&=\begin{cases}
		0 & \text{if }
		\displaystyle\rho_{test}-\sum_{j=i+1}^{m}\sum_{c=0}^{N_c}f_{j,c}(x-e_j\delta_t)\frac{n_l(x-{e}_{j}\delta _{t})}{n_l(x)}>1 \\
		f_{i,c}(x-e_i\delta_t) \phantom{ \ + f_{i-1,c}(x-e_i\delta_t)} & \text{else}
		\end{cases} \\
		f_{i-1,c}^{*}(x-e_i\delta_t)&=\begin{cases}
		f_{i,c}(x-e_i\delta_t)+ f_{i-1,c}(x-e_i\delta_t) & \text{if }
		\displaystyle\rho_{test}-\sum_{j=i+1}^{m}\sum_{c=0}^{N_c}f_{j,c}(x-e_j\delta_t)\frac{n_l(x-{e}_{j}\delta _{t})}{n_l(x)}>1 \\
		f_{i-1,c}(x-e_i\delta_t) & \text{else}
		\end{cases}
		\end{cases} \label{eq_virt_bound_mp}\\
		{\rho_{test}}(x)=&\displaystyle \sum_{i=0}^{m} \sum_{c=0}^{N_c} f_{i,c}(x-e_i\delta_t)\frac{n_l(x-{e}_{i}\delta _{t})}{n_l(x)}. \label{eq_rho_test_mp}
		\end{align}
	\end{widetext}

	\section{Numerical Validating Simulations}
	
	In all the following simulations, we take a space step of $5.5$ metres and a time step of $1$ second. This leads to a speed step of almost $20$ kilometres per hour. Moreover, we used a $D1Q6$ schema, consequently the maximum speed is close to $100$ kilometres per hour. But for the sake of generality, the following results will be presented in their non-dimensional form, i.e. expressed in space or time step units.
	
	All the following simulations are obtained with a relaxation frequency of $0.9$, at the exception of the simulations linked to the fig.~\ref{fig_diag_fund_omega} where the value of the relaxation time is varying.
	
	\subsection{Fundamental Diagrams}		
\begin{figure}[h!]
		\def\svgwidth{0.55\linewidth}
		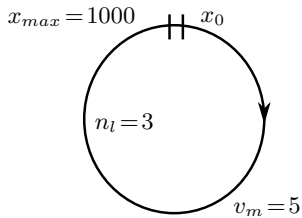
		\vspace{-20pt}
		\caption{Schematic of a ring road.}
		\label{ring_road_cars}
	\end{figure}

\begin{figure}
		\includegraphics{\PathS 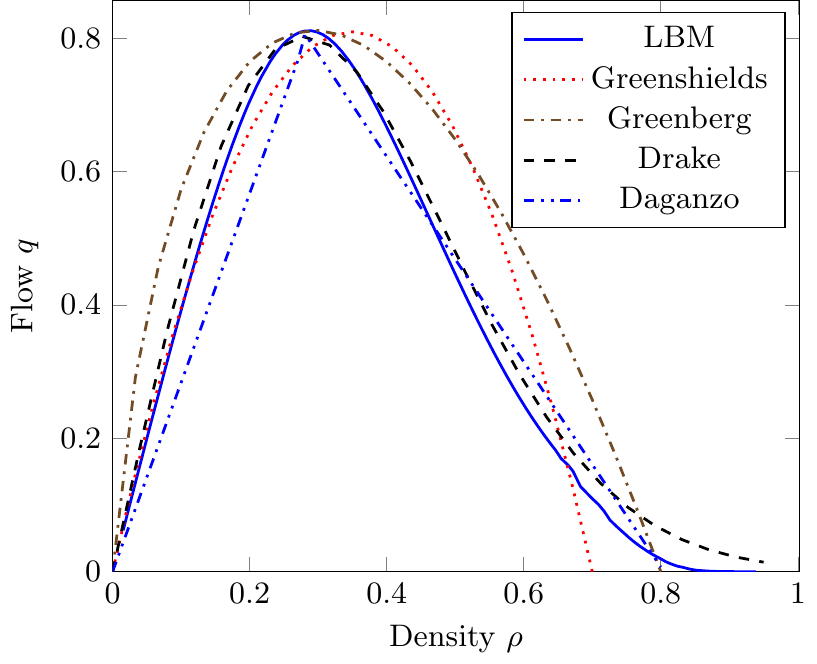}
		\caption{Fundamental diagram of some macroscopic models}
		\label{fig_diag_fund_ref}
	\end{figure}

	In order to evaluate the behaviour of the suggested model, the study of density-flow fundamental diagram is performed, which allows estimating the relationship between the flow and the density~\cite{Kerner_2002, Ngoduy_2011}.
	The study of a ring road, of $1000$ cell length, simulated for different average density. Each simulation starts with a density spatially varying around the average density by following a random noise of $10$\%. The road has a speed restriction of $5$ cells per unit of time (see fig.~\ref{ring_road_cars}). The relaxation time is set to $0.9$. After $2000$ time steps, the averaging over time and space is made to obtain the fig.~\ref{fig_diag_fund_ref}. The resulting curve is compared to usual macroscopic models such as the Greenshield~\cite{Greenshields_1935}, Greenberg~\cite{Greenberg_1959}, Drake~\cite{Drake_1967} or Daganzo~\cite{Daganzo_1997} models. 
	
	The fig.~\ref{fig_diag_fund_ref} shows the ability of the Lattice-Boltzmann method to simulate the various traffic situations with good accuracy. The results are very close to those described by the Drake model~\cite{Drake_1967}.
	
\begin{figure}
		\includegraphics{\PathS 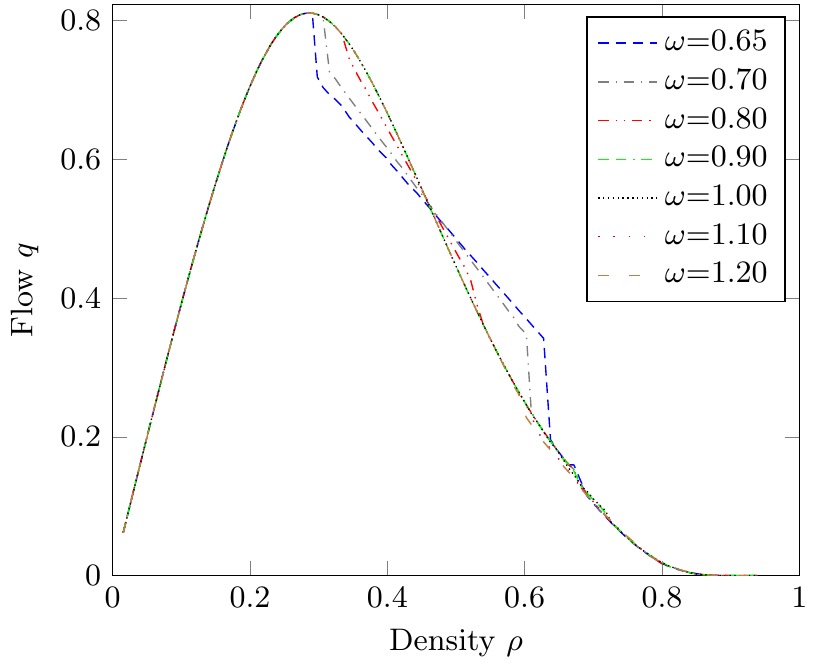}
		\caption{Flow-density relationship for different relaxation time}
		\label{fig_diag_fund_omega}
	\end{figure}

	Moreover, the effect of the relaxation time on the fundamental diagram is major. This effect is represented on the fig.~\ref{fig_diag_fund_omega}. The relaxation time interval in which the numerical schema remains stable is directly linked to the equilibrium distribution function suggested. And the fact that its effect is only noticeable for congested-flow can be interpreted as the too slow ability of the drivers to change (or reach) their desire speed. In other words, they have a speed incompatible with the current density of traffic.
	
	For values of the relaxation time higher than $1.20$ and lower than $0.65$, the numerical model becomes unstable for high densities.

	\subsection{Road Merging}
\begin{figure}[h!]
		\def\svgwidth{0.99\linewidth}
		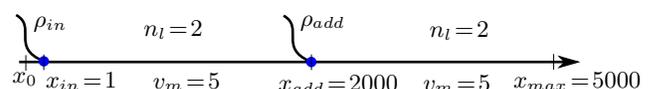
		\caption{Schematic of merging roads.}
		\label{road_inject}
	\end{figure}

	Changes of the traffic conditions can be caused by road merging. To study the effects of road merging on traffic conditions, the simulation of $2$-lane highway is considered. This road is constituted of $5000$ cells and the speed limit is $5$ cells per unit of time. The complete highway is empty at the time $0$ of the simulation. At the beginning of the road, the density is set to a constant while $2000$ cells after a road merging add another constant density except if the sum is higher than one (see fig.~\ref{road_inject}).
	
\begin{figure}
		\pgfplotsset{width=0.99\linewidth}
		\includegraphics{\PathS 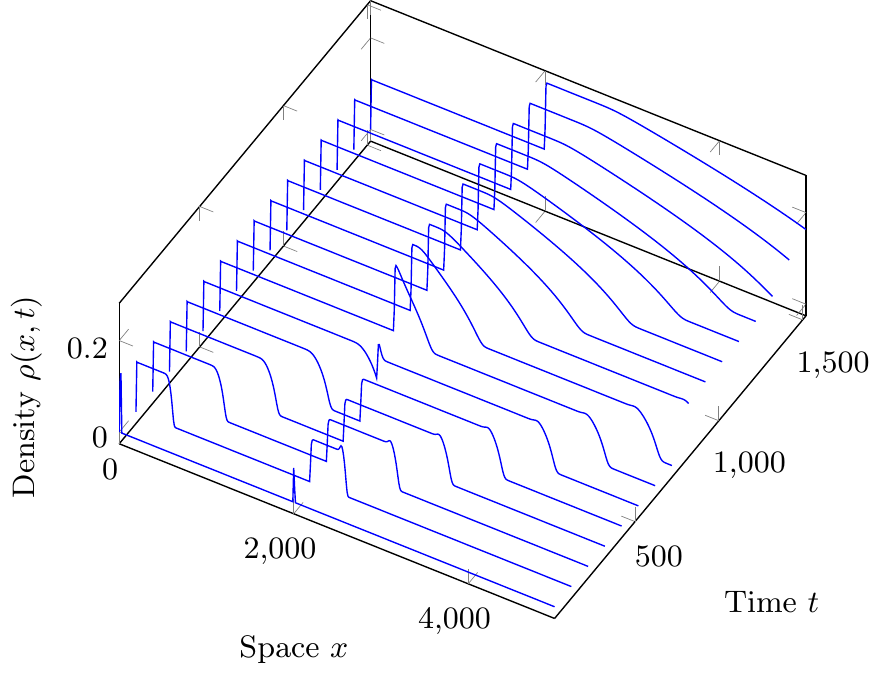}
		\caption{Modelling of road merging in free-flow traffic conditions.}
		\label{fig_injec_nojam}
	\end{figure}

	The fig.~\ref{fig_injec_nojam} illustrates the dynamic of the densities when the two flows merge. The density at the edge of the simulation is set to $0.11$ and the incident density is set to $0.15$. These two densities and there sum are in the free-flow conditions. Therefore, no jam is observed and after the merge, when the flows joined the density is simply the sum of the incoming densities.
	
\begin{figure}
		\includegraphics{\PathS 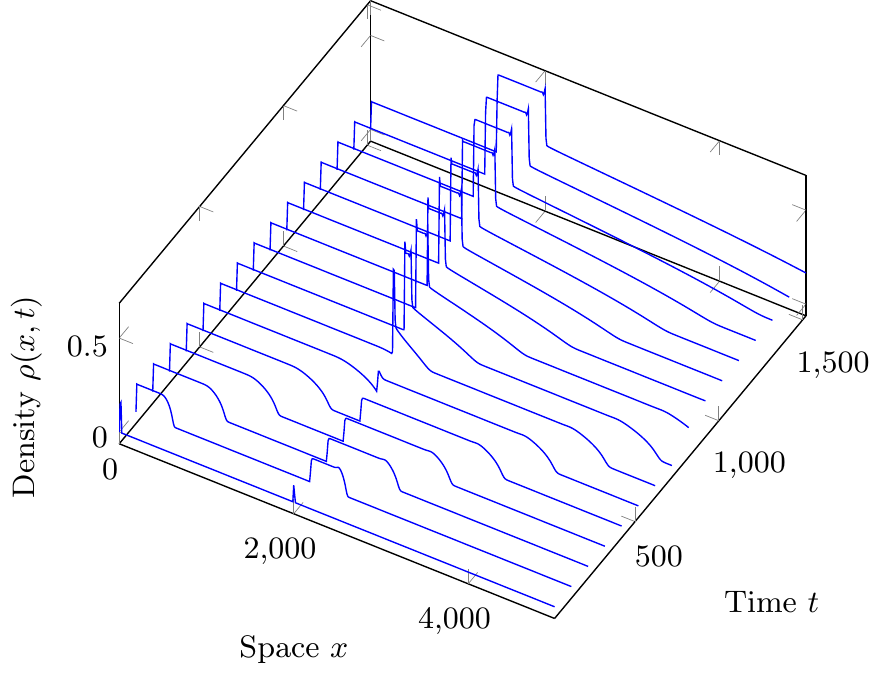}
		\caption{Modelling of road merging in congested-flow traffic conditions.}
		\label{fig_injec_jam}
	\end{figure}

	{Per contra}, the fig.~\ref{fig_injec_jam} is obtained with an initial density of $0.15$ at the beginning of the road and an incident density of $0.20$. These two ones are still in the free-flow domain but not their sum, therefore as expected the fig.~\ref{fig_injec_nojam} shows a slowing down waves (characterised by an increase of the density) streaming backward from the merging point.
	
	\subsection{Number of Lane}
\begin{figure}[h!]
		\def\svgwidth{0.99\linewidth}
		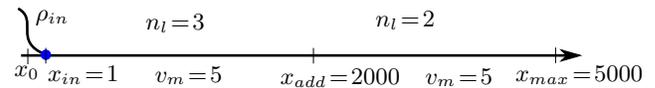
		\caption{Schematic of a road with number of lane change.}
		\label{road_lane}
	\end{figure}

	To evaluate the effect of a change of lane number on traffic conditions; the study of a road of $5000$ sites is proposed. It starts with $3$ lanes and a constant density, while a reduction to $2$ lanes is located at site number $2500$. The speed limit is set to $5$ cells per unit of time, see fig.~\ref{road_lane}.		
	
	\begin{figure}[h!]
		\includegraphics{\PathS 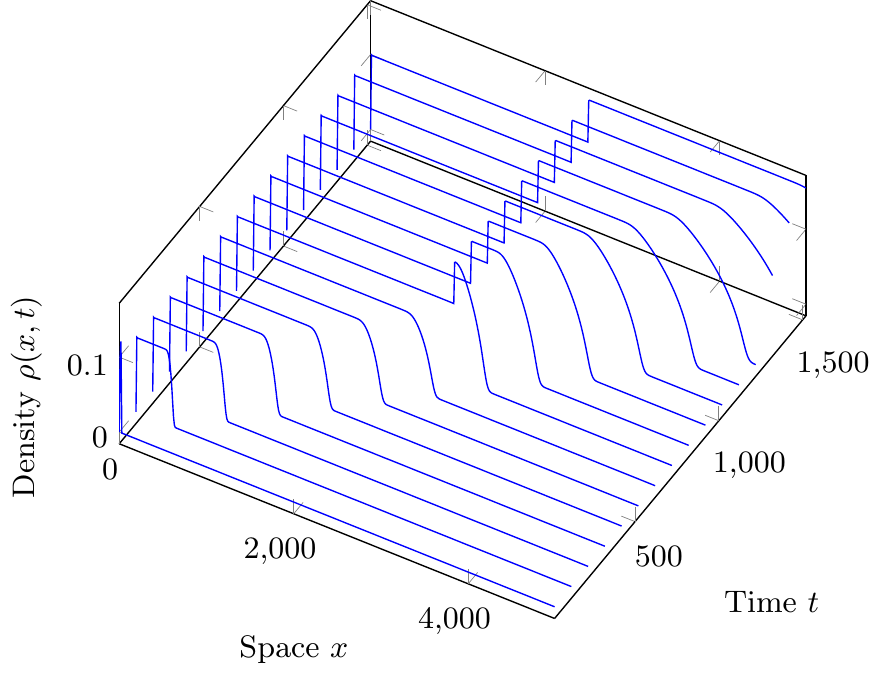}
		\caption{Density for reduction of lanes in free-flow conditions.}
		\label{fig_lane_nojam}
	\end{figure}
	
	\begin{figure}[h!]
		\includegraphics{\PathS 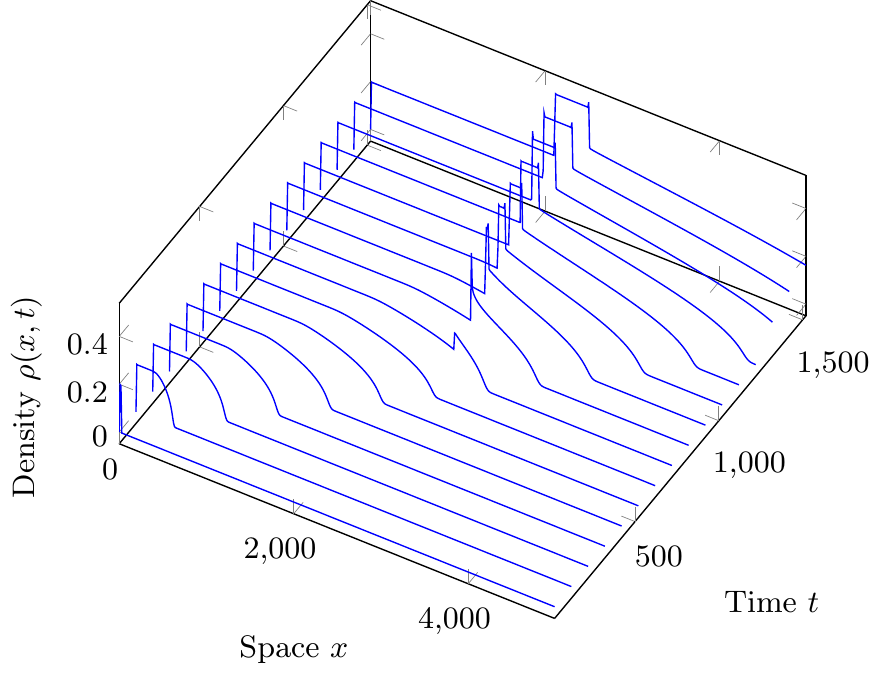}
		\caption{Density for a reduction of lanes in congested-flow traffic conditions.}
		\label{fig_lane_jam}
	\end{figure}
	
	The fig.~\ref{fig_lane_nojam} represents density evolution in time and space for a simulation with the entrance density of $0.10$. 
	This situation leads to a density increase after the reduction of lanes, which remains in the free-flow domain.
	
	The fig.~\ref{fig_lane_jam}, has for incoming density $0.20$, which means an increase of its density will bring it into the congested-flow domain and might create important interactions. On the fig.~\ref{fig_lane_jam}, the entrance in the congested-flow domain leads to a slowing-down situation waving backward.

	\subsection{Speed Limit Change}
\begin{figure}[h!]
		\def\svgwidth{0.99\linewidth}
		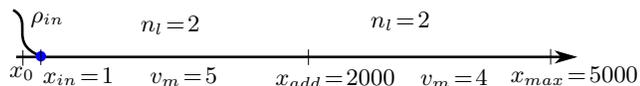
		\caption{Schematic of a road with a change of speed limit.}
		\label{road_speed}
	\end{figure}

	The last external source of modification for traffic conditions we studied, is the change of speed limit. The same length of the road is used than the previous numerical investigations for a road of $2$ lanes and the same relaxation frequency. A reduction of speed limits from $5$ to $4$ cells per unit of time is imposed at site number $2500$, see fig.~\ref{road_speed}.
	
\begin{figure}
		\includegraphics{\PathS 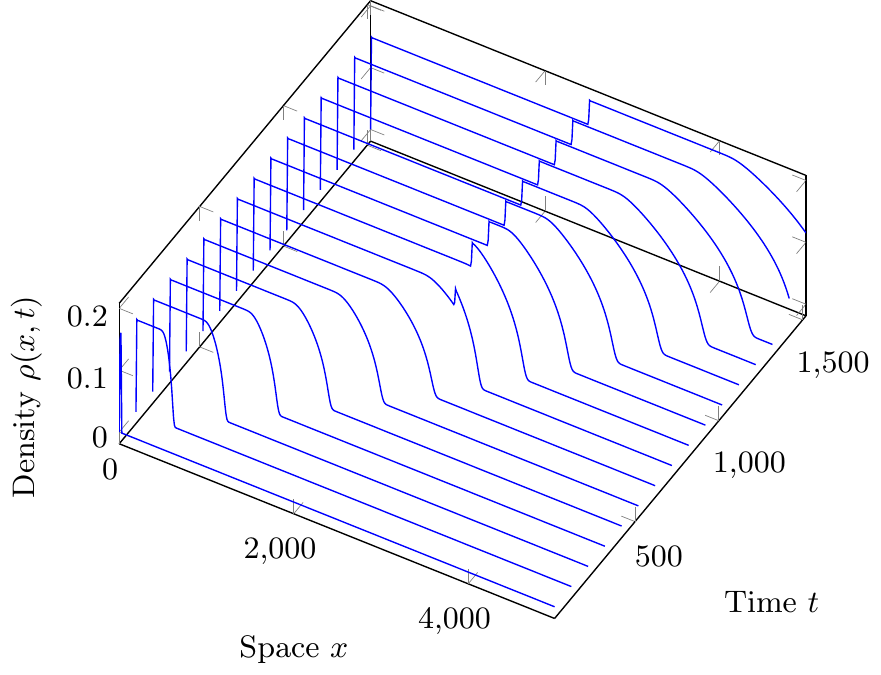}
		\caption{Modelling of a road containing a reduction of speed restriction under free-flow traffic conditions.}
		\label{fig_speed_nojam}
	\end{figure}

	With a density of $0.15$ at the entrance of the road, the speed limit reduction has the same effect as the reduction of lane number: an increase of the density forward to the change. The fig.~\ref{fig_speed_nojam} represents this situation with the increased density still in the free-flow domain. 
	
\begin{figure}
		\includegraphics{\PathS 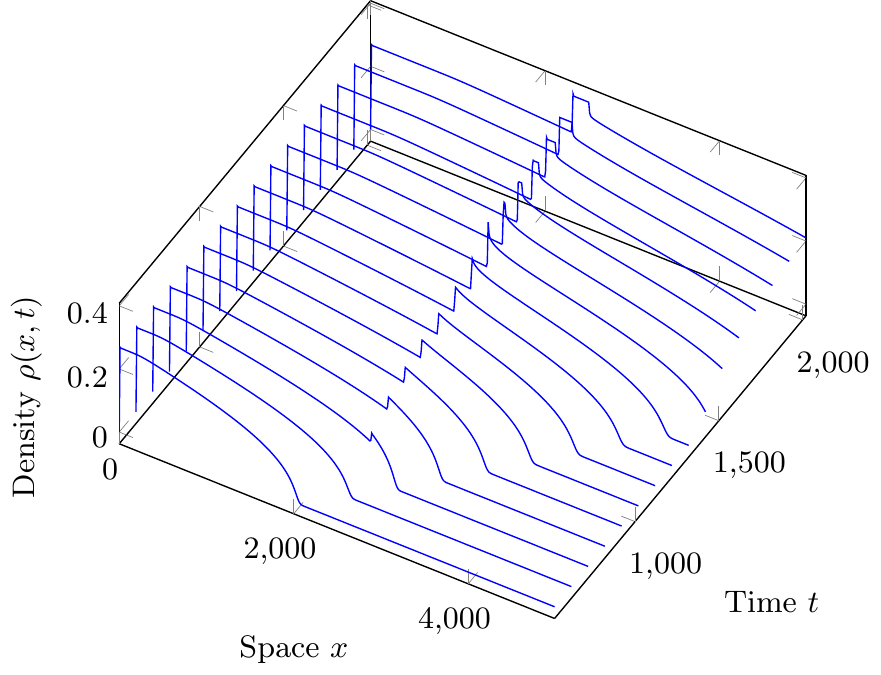}
		\caption{Modelling of a road containing a reduction of speed restriction under congested-flow traffic conditions.}
		\label{fig_speed_jam}
	\end{figure}

	For the fig.~\ref{fig_speed_jam}, the entrance density set at $0.26$ leads, after the reduction of speed limits, to a congested flow; the same consequences as the previous congested cases, i.e. a slowing down flow. 
	
	\subsection{Truck Concentration}
\begin{figure}[h!]
		\def\svgwidth{0.5\linewidth}
		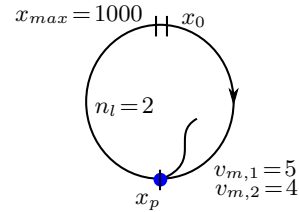
		\caption{Schematic of a ring road with an injection point.}
		\label{ring_road_truck}
	\end{figure}

	To evaluate the effects of heterogeneous multi-class traffic, a ring road is studied. 
	The ring road is made of $1000$ cells, $2$ lanes and an injection point is positioned at $x_p$ (the exact index of $x_p$ does not matter because of the circularity of the numerical model), as the fig.~\ref{ring_road_truck}. The injection point imposes a global density $\rho_p$ distributed in two classes, through a coefficient $\alpha$. Thus, $\rho_1(x_p)= \alpha\rho_p$ and $\rho_2(x_p)= (1-\alpha)\rho_p$.
	
\begin{figure}
		\pgfplotsset{width=0.99\linewidth}
		\includegraphics{\PathS 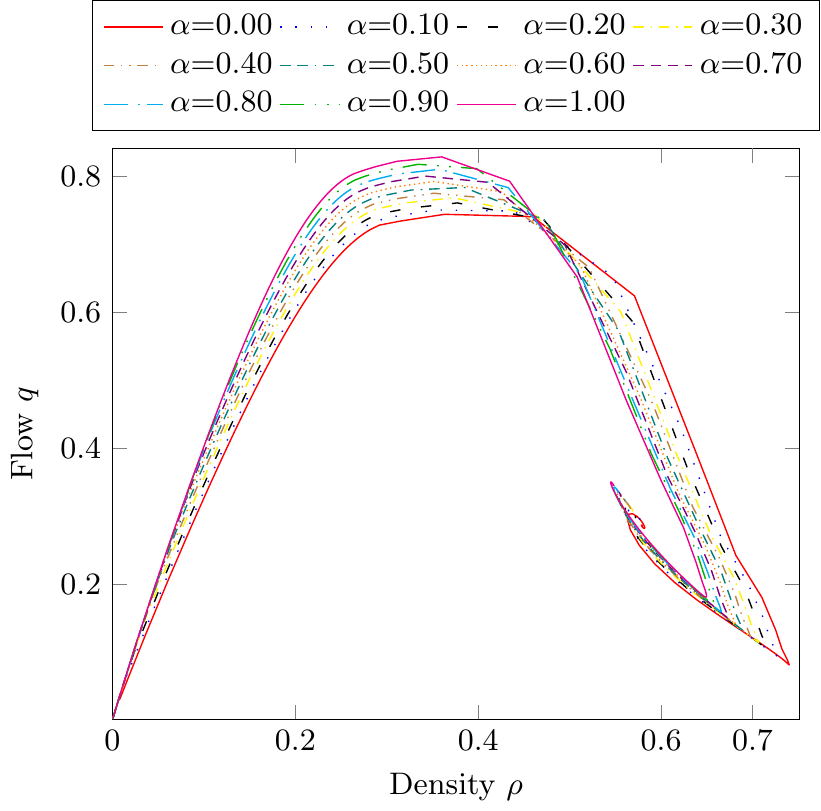}
		\caption{Fundamental diagram: flow-density relationship for different lorry concentration.}
		\label{fig_diag_fund_truck}
	\end{figure}

	The fig.~\ref{fig_diag_fund_truck} shows the influence of class of slower vehicles (with index $2$), with speed limit of $4$ cells per unit of time (to simulate the heavy weighted machines), on a faster class (with index $1$) having a speed limit of $5$ cells per unit of time (to model the personal cars). The density $\rho_p$ is set to $0.725$, in order to run through almost all the traffic domains. The fig.~\ref{fig_diag_fund_truck} shows the fundamental diagram of one point $200$ cells after $x_p$, without time nor space averaging. This lack of averaging allows visualising the hysteresis phenomenon captured by the LBM, and can be understood through the third phase in fundamental diagrams~\cite{Kerner_2002}. 
	
	Moreover, the latter figure highlights that the increase of slower vehicles makes the jam and congested situations appear earlier. Plus, the maximum possible flow is lower. This is even more drastic when re-dimensioning the fundamental diagram. Indeed, if the slower vehicles are lorries since they are more represented through the eq.~(\ref{eq_dim}).

	\section{Discussions \& Conclusions}
	The suggested method allows to reproduce macroscopic situations from a mesoscopic formulation through a Boltzmann-like equation.
	The present lattice-Boltzmann method is capable of modelling the ideal traffic behaviour described in the literature (such as Drake model), and the basic psychological behaviour of the drivers through the relaxation time parameter. 
	The improved formulation allows simulating various road situations, unprecedented with such method. The method is able to deal with road merging, change of lane numbers or speed limits in both free-flow and congested-flow conditions in accordance with the macroscopic previsions.
	The influence of a multi-class traffic found here is close to those presented by some authors~\cite{Ez-Zahraouy_2004}.
	
	To conclude, the lattice Boltzmann method is an efficient numerical method to overcome the integro-differential difficulties introduced by statistical models and might be a practical manner to solve numerically the Prigogine-Boltzmann like equation. This remains a good compromise between, on the one hand, the level of detail but time consuming provided by the microscopic description and, on the other hand, the loss of information but
	the much faster computation yield by the macroscopic description. Moreover, the macroscopic results are bounded to the choice of the equilibrium density function that can be tuned to reproduce various effects and models.
	
	The present results prove for the first time the capacity of this method to solve heterogeneous multi-class traffic flows with precision. And suggested formulations give easy treatment to deal with numerous road situations.
	
	This work could be extended by adding of new parameters involved to model certain traffic situation. An example could be to incorporate the psychological behaviour of drivers. Other aspects such as the sinuosity of the road or the traffic pressure in multi-class flows could also be interesting. Investigations on the equilibrium distribution functions and their justification, would be a major work.
	
	\begin{acknowledgments}
		This work has been partially funded by the French National Research Agency via the LBSMI project ANR-15-CE19-0002.
		The authors would like to thank Solenn \textsc{Bardaud} and Elias \textsc{Abou Rachid} for their participation to numerical implementations.
	\end{acknowledgments}
	
	
	\bibliographystyle{apsrev4-1}
	\bibliography{\PathS /LBM_traffic-multiclass}
	
\end{document}